\documentclass[aps,preprint,prb,showpacs,amsmath,floatfix,superscriptaddress]{revtex4-1}

\usepackage{amsmath}
\usepackage{amssymb}
\usepackage{graphicx}
\usepackage{bbold}
\usepackage{times}
\usepackage{hyperref}
\usepackage{dsfont}
\usepackage{color}
\begin{document}

\title{Symmetry between repulsive and attractive interactions in driven-dissipative Bose-Hubbard systems}

\author{Adil A. Gangat}
\affiliation{Department of Physics, National Taiwan University, Taipei 10607, Taiwan}
\author{Ian P. McCulloch}
\affiliation{ARC Centre of Excellence for Engineered Quantum Systems, School of Mathematics and Physics, The University of Queensland, St. Lucia, QLD 4072, Australia}
\author{Ying-Jer Kao}
\affiliation{Department of Physics, National Taiwan University, Taipei 10607, Taiwan}
\affiliation{National Center of Theoretical Sciences, National Tsinghua University, Hsinchu 30013, Taiwan}
\affiliation{yjkao@phys.ntu.edu.tw}

\begin{abstract}
The driven-dissipative Bose-Hubbard model can be experimentally realized with either negative or positive onsite detunings, inter-site hopping energies, and onsite interaction energies.  Here we use one-dimensional matrix product density operators to perform a fully quantum investigation of the dependence of the non-equilibrium steady states of this model on the signs of these parameters.  Due to a symmetry in the Lindblad master equation, we find that simultaneously changing the sign of the interaction energies, hopping energies, and chemical potentials leaves the local boson number distribution and inter-site number correlations invariant, and the steady-state complex conjugated.  This shows that all driven-dissipative phenomena of interacting bosons described by the Lindblad master equation, such as ``fermionization" and ``superbunching", can equivalently occur with attractive or repulsive interactions.
\end{abstract}

\maketitle

\section*{Introduction}
The non-equilibrium behaviour of Bose-Hubbard systems has received considerable theoretical attention recently \cite{carusotto2009fermionized, hartmann2010polariton, grujic2013repulsively, le2014bose, biella2015photon, Weimer2015-is, foss2016emergent, debnath2017nonequilibrium}.  However, to our knowledge the dependence of the non-equilibrium physics of the Bose-Hubbard model (BHM) on the \textit{signs} of the hopping and interaction energies has yet to be explored.  In superconducting circuits, which are a natural setting for studying the non-equilibrium physics of driven-dissipative many-body systems \cite{Houck2012-cu,schmidt2013circuit,hur2015many}, strong interactions are more accessible with attractive interaction energies than with repulsive interaction energies \cite{bourassa2012josephson,hacohen2015cooling}.  On the other hand, theoretical studies of the driven-dissipative BHM (DDBHM) have focused on the case of repulsive interactions.  Finding a theoretical link between the attractive and repulsive interaction regimes of the DDBHM would therefore be of practical experimental benefit.


In this work we point out a symmetry in the Lindbladian equation of motion for the DDBHM that implies that the driven-dissipative physics of repulsive interactions can be replicated with attractive interactions, irrespective of the magnitude of the interaction strength.  To illustrate this, we employ a fully quantum (i.e. non-mean-field) numerical treatment of a DDBHM trimer.  We show that simultaneously changing the signs of the interaction, hopping, and detuning while keeping their magnitudes fixed changes the NESS but does not change the three-site number correlator nor the statistics of the on-site boson number.  We also demonstrate that this observable symmetry persists even in the presence of strong disorder in all of the sign-flipped parameters.  This symmetry can be experimentally tested with existing superconducting circuit technology, which has the potential to realize the BHM such that the chemical potential, on-site interaction energy, and inter-site hopping energy are all tunable \textit{in situ} (within a limited range) in both magnitude and sign \cite{hacohen2015cooling,deng2015sitewise,deng2016superconducting}.

\section*{Model}
We  investigate the open boundary dissipative Bose-Hubbard chain under homogeneous coherent driving in a frame rotating at the drive frequency.  With on-site dissipation to a Markovian bath, the effective equation of motion (EOM) (see Appendix for derivation) is given by the following Lindblad master equation ($\hbar=1$): 
\begin{eqnarray}
\frac{d}{dt}{\rho}&=&\mathcal{{L}}{\rho}=-i[{H},{\rho}] \nonumber\\ &~&~~~~~~~~~~~~+ \gamma\sum_l \frac{1}{2}(2b_l{\rho}b_l^{\dagger}-b_l^{\dagger}b_l{\rho}-{\rho}b_l^{\dagger}b_l),\label{EOM}\\
{H}&=& \sum_l \Delta_{l}b_l^{\dagger}b_l - \sum_l J_{l,l+1}(b_l^{\dagger}b_{l+1} + b_lb_{l+1}^{\dagger}) \nonumber \\  &~&~~~~~~+ \sum_l \frac{U_l}{2}b_l^{\dagger}b_l^{\dagger}b_lb_l + \Omega\sum_l(b_l^{\dagger}+b_l),
\end{eqnarray}
where $J_{l,l+1}$ denotes the hopping amplitude between the $l$th and ($l+1$)th site, $U_l$ denotes the boson interaction energy on the $l$th site, $\gamma$ is the local dissipation rate, $\Omega$ denotes the drive amplitude (assumed real), and $\Delta_{l}=\omega_l-\omega_d$, which plays the role of a chemical potential, is the site-dependent drive detuning when $\omega_{l}$ is the bare frequency of the $l$th site and $\omega_{d}$ is the drive frequency.

The NESS of the DDBHM, denoted ${\rho}_{\infty}$, is defined as  the fixed point of the evolution given by equation (\ref{EOM}),  $\frac{d}{dt}{\rho}_{\infty}=0$.  We observe that the EOM for ${\rho}$ given by equation (\ref{EOM}) is the same as the EOM for ${\rho}^*$ if the Hamiltonian is negated (${H}\rightarrow -{H}$).  Therefore the NESS attained by evolving with ${H}$ is equal to the complex conjugate of the NESS attained by evolving with $-{H}$. However, the transformation ${\rho}\rightarrow{\rho}^*$ does not change the observable statistics of the state. The observables of the NESS are therefore invariant under negation of the Hamiltonian.  We note that this symmetry applies not just to the DDBHM, but to any model described by the Lindblad master equation where the dissipation operators are invariant under complex conjugation.

For the DDBHM there is a further simplification of the symmetry.  The transformation ${H}\rightarrow -{H}$ entails $\Omega\rightarrow -\Omega$, which is equivalent to $b_l \rightarrow -b_l$.  However, $b_l \rightarrow -b_l$ itself does not change the boson number statistics.  To see this, note that if $b \rightarrow -b$, then 
\begin{equation}
|n\rangle\langle n| = (b^{\dagger})^n|0\rangle\langle0|(b)^n \rightarrow (-b^{\dagger})^n|0\rangle\langle0|(-b)^n = |n\rangle\langle n|.  
\end{equation}
We therefore conclude that $\Omega\rightarrow -\Omega$ is unnecessary to preserve the boson number statistics in the NESS; the invariance only requires $U_l\rightarrow-U_l$, $J_{l,l+1}\rightarrow - J_{l,l+1}$, and $\Delta_l\rightarrow-\Delta_l$. 

\section*{Numerical Simulation}
The numerical simulation is performed by employing a  matrix product density operator (MPDO) representation of ${\rho}$ \cite{Zwolak2004-tz, Verstraete2004-hc}, which amounts to a quantum mechanical treatment characterized by a refinement parameter $\chi$ that designates the maximum size of the tensors that represent each site, and therefore the maximum amount of total correlations (classical plus quantum) between bipartitions of the chain that can be captured by the MPDO.  Linking each site tensor with its neighbor in the MPDO is a diagonal matrix of $\chi$ ``singular values" that represents these correlations.

In the MPDO picture the system density matrix ${\rho}$ becomes a vector, denoted $|{\rho}\rangle$, and the superoperator $\mathcal{{L}}$ becomes a regular operator $\mathcal{{L}}_\sharp$ such that $\langle {\rho}|\mathcal{{L}}_\sharp |{\rho}\rangle = 0$ at the NESS. To obtain an approximation for ${\rho}_{\infty}$ under a given set of system parameters $U_l$, $J_l$, $\Delta_l$, $\Omega$, and $\gamma$, we first use the hybrid evolution method of Ref. \cite{gangat2016steady} to evolve the MPDO representation of a random initial state ${\rho}$ under a desired choice of parameters until convergence in achieved.  We then sweep the value of $\Omega$ in increments, converging the MPDO with real time evolution at each increment.  Convergence is considered complete when $\langle \mathcal{{L}}_\sharp \rangle \lesssim 10^{-3}$ and the singular values between the first two sites of the MPDO are converged on a logarithmic scale.  We find that $\chi=15$ and a timestep size of $10^{-1}$ is sufficient to achieve this for all of the cases that we consider.  We verify uniqueness of the NESS by performing the sweep of $\Omega$ in both directions.  We truncate the Hilbert space on each site at four quanta, and always choose $\gamma=1$.

\begin{table}
\begin{center}
\begin{tabular}{l|rrr|rrr}

&\multicolumn{3}{l} {(a) Uniform}  & \multicolumn{3}{|l} {(b) Disordered}  \\
\hline
Case & $J$ & $U$ & $\Delta$  & $J_{1,2}$, $J_{2,3}$ & $U_1$, $U_2$, $U_3 $  & $\Delta_1$, $\Delta_2$, $\Delta_3$\\
\hline\hline
1 &$\pm1$ & $\pm10$ & $\pm 1$   & $\pm 1, \mp 3$ & $\pm 8, 0, \mp 10$ & $0, \mp 10,  \pm1$\\
2 & $1$ & $\pm 10$& $\pm 1$ &  $1, -3$ & $\pm 8, 0,  \mp 10$ & $0, \mp 10, \pm 1$
\end{tabular}
\end{center}
\caption{\label{tab1} \textbf{Simulation parameters for the DDBHM trimer.}     Case 1 corresponds to a number-conserving transformation in which the signs of all the parameters are flipped simultaneously. Case 2 corresponds to a transformation in which the sign of the hopping energy ($J$) is kept fixed while the sign of the interaction strength ($U$) and detuning ($\Delta$) are changed. $\gamma=1$ for all cases. \textbf{(a)} Uniform trimer, \textbf{(b)} Disordered trimer with non-uniform parameters.  }
\end{table}

\section*{Results}

\subsection{Uniform trimer}
To test the arguments set forth above, we perform numerical investigations on a DDBHM trimer system.  We first test the boson number symmetry when the parameters are uniform across the trimer.  We specifically look at two cases:  Case 1 examines the change in the NESS under the number-conserving transformation argued above (the hopping energy $J$, the interaction strength $U $, and the  detuning $\Delta$ all change signs simultaneously);  Case 2 examines the change in the NESS under a transformation that is different from the number-conserving transformation discussed in the previous section: the sign of  $J$ is kept fixed while the sign of  $U$ and  $\Delta$ are changed. The simulation parameters are summarized in Table~\ref{tab1}.
Both cases are examined at several different values of the drive strength $\Omega$.  At each value of $\Omega$ we denote the NESS for upper and lower sign choices by $|{\rho}^{(+)}_\infty\rangle$ and $|{\rho}^{(-)}_\infty\rangle$, respectively.

\begin{figure}
\centerline{
\includegraphics[scale=0.6]{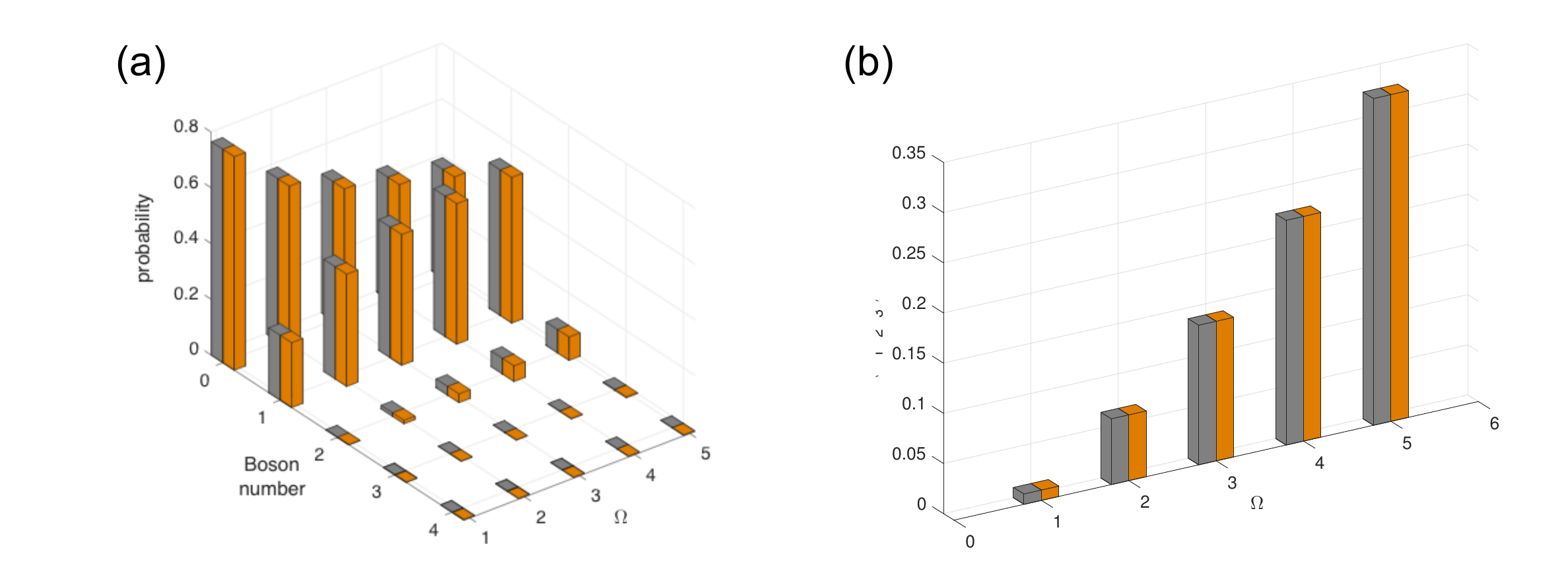}
}
\caption{\label{fig_uniform1}\textbf{ Boson number statistics for site 1 and three-site correlators under a number-conserving transformation in a uniform trimer. }  Gray bars correspond to the lower sign choice in the parameters listed in Table~\ref{tab1}(a), Case 1; and orange bars correspond to the upper sign choice.  \textbf{(a)} Boson number statistics as a function of drive strength $\Omega$. \textbf{(b)} Three-site correlator  as a function of drive strength. Although the NESS changes between the two different sign choices at each drive strength, the number statistics  and the correlator remain the same. }

\end{figure}

First we examine the parameter sets for Case 1  in Table~\ref{tab1}(a).
In accordance with the boson number symmetry argued earlier, here we find that at every value of $\Omega$ the local and non-local observables $n_1$ and $n_1n_2n_3$ are invariant in the NESS under the collective sign change, as shown in Fig.~\ref{fig_uniform1}.  More precisely, we see that the full statistical distribution of $n_1$ is the same.  

\begin{figure}
\centerline{
\includegraphics[width=0.5\textwidth]{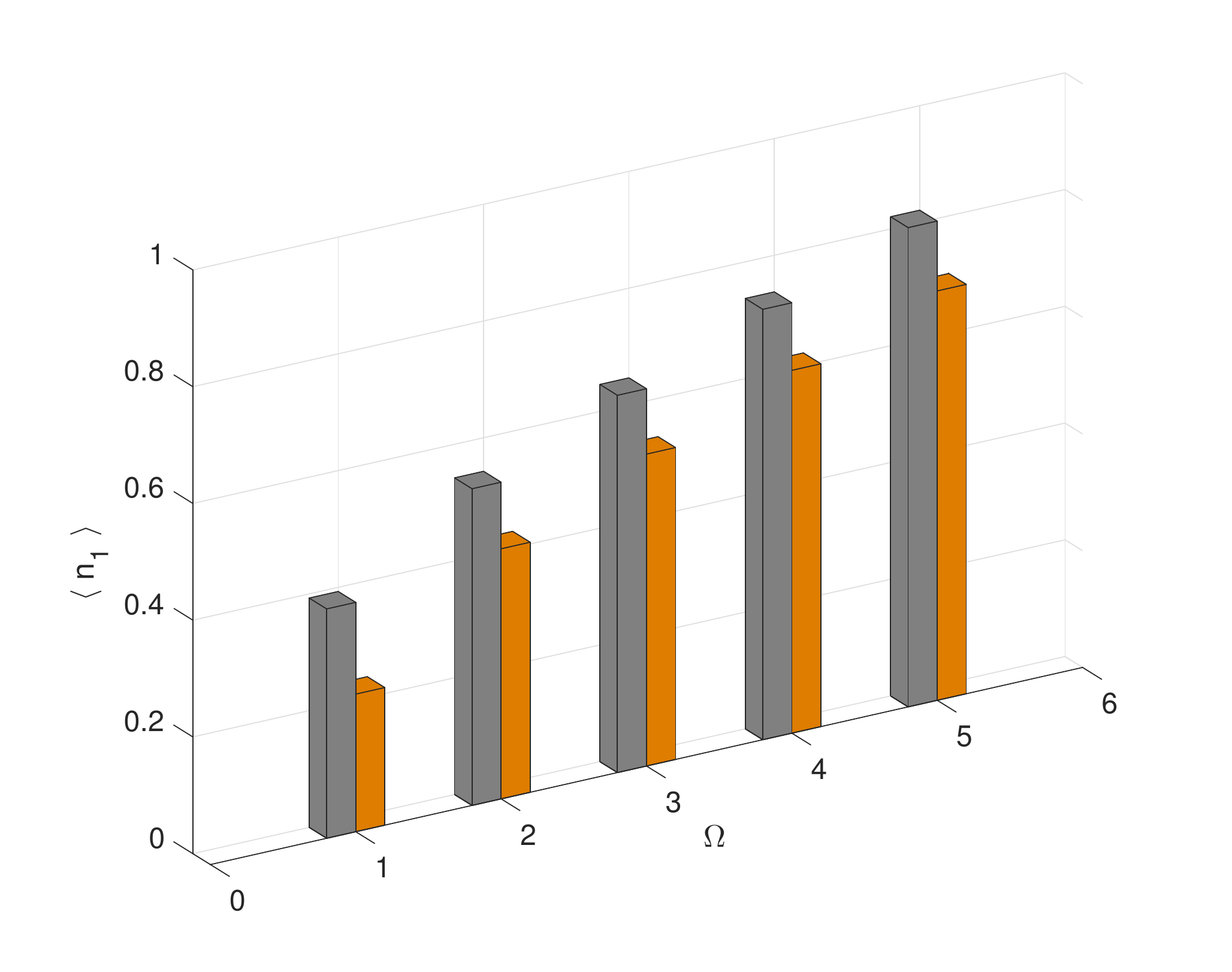}
}
\caption{\label{fig_uniform2}\textbf{Boson number expectation value on site 1 as function of varying drive strength ($\Omega$) for a uniform trimer.} Gray bars correspond to the lower sign choice in the parameters listed in Table~\ref{tab1}(a), Case 2; and orange bars correspond to the upper sign choice. The transformation from the lower sign choice to the upper sign choice is not number-conserving and the boson number expectation value is not invariant.  }

\end{figure}

Next we examine Case 2 where the sign of  $J$ is kept fixed while the sign of  $U$ and  $\Delta$ are changed. This is not a number-conserving transformation and we do not expect the observables will remain the same after the transform. 
In Fig.~\ref{fig_uniform2}, we plot the expectation value of the observable $n_1$ for both $|{\rho}^{(+)}_\infty\rangle$ and $|{\rho}^{(-)}_\infty\rangle$. We find that $\langle n_1\rangle$ differs between $|{\rho}^{(+)}_\infty\rangle$ and $|{\rho}^{(-)}_\infty\rangle$ at each value of $\Omega$, and therefore conclude that the number statistics is not invariant under only $U_l\rightarrow-U_l$ and $\Delta_l\rightarrow-\Delta_l$. This case is similar to the interaction  sign change  in the equilibrium BHM , where the hopping energy remains fixed, and the equilibrium phase changes.

\subsection{Disordered trimer} 
To further demonstrate that the invariance is very general, we now test the boson number symmetry in the presence of strong disorder. As before, $|{\rho}^{(+)}_\infty\rangle$ and $|{\rho}^{(-)}_\infty\rangle$ respectively denote the upper and lower sign choices of the parameters.  We consider two specific cases analogous to those for the uniform trimer. In Case 1, the change in the NESS is examined when the  hopping energy, interaction strength, and detuning all change sign; In Case 2, the sign of the hopping energy is kept fixed while the signs of the interaction strength and detuning are changed.

\begin{figure}
\centerline{
\includegraphics[width=\textwidth]{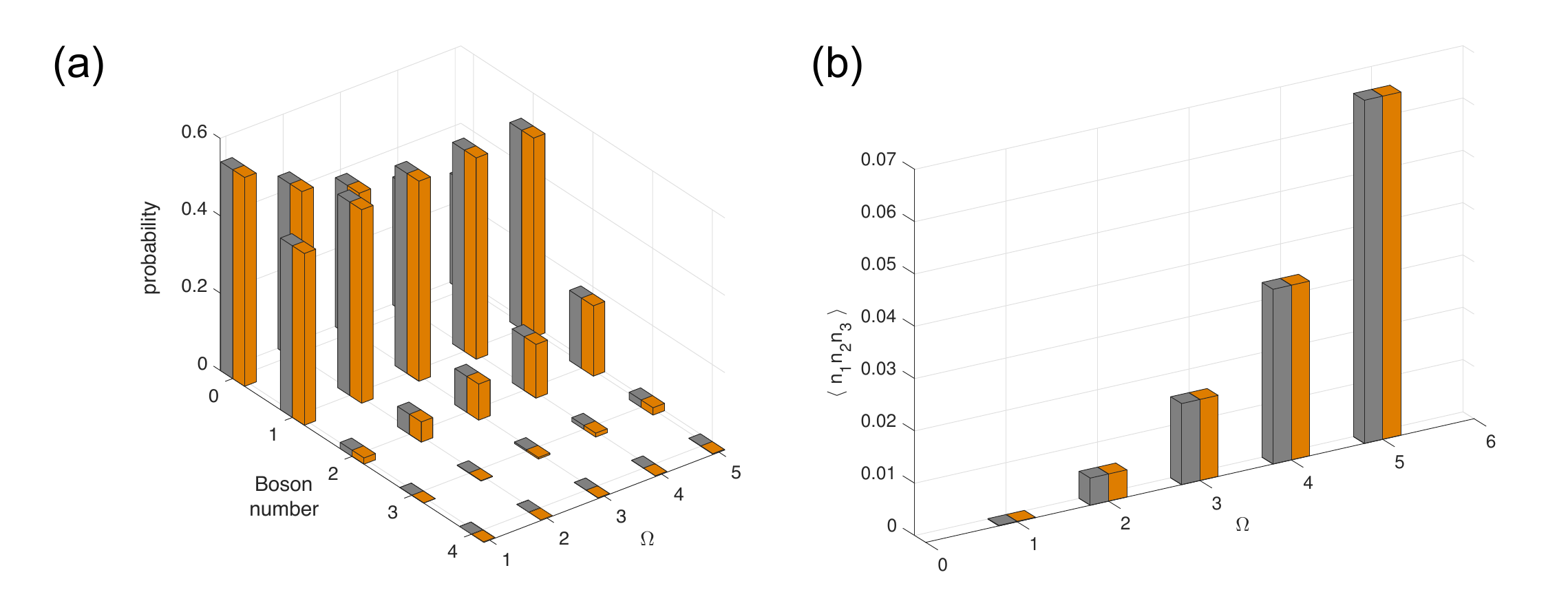}
}
\caption{\label{fig_disordered1}\textbf{Boson number statistics for site 1 and three-site correlators under a number-conserving transformation in a disordered trimer. }  Gray bars correspond to the lower sign choice in the parameters listed in Table~\ref{tab1}(b), Case 1; and orange bars correspond to the upper sign choice.  \textbf{(a)} Boson number statistics as a function of drive strength $\Omega$. \textbf{(b)} Three-site correlator  as a function of drive strength. Although the NESS changes between the two different sign choices at each drive strength, the number statistics  and the correlator remain the same even in the presence of strong disorder. }
\end{figure}

We first examine parameters for Case 1 as listed in Table~\ref{tab1}(b).
In this case the upper sign choice and lower sign choice of the parameters are related by the boson number symmetry transformation.  Consequently, Fig.~ \ref{fig_disordered1} reveals that the local and non-local observables $n_1$ and $n_1n_2n_3$ are the same between $|{\rho}^{(+)}_\infty\rangle$ and $|{\rho}^{(-)}_\infty\rangle$ at any given $\Omega$. We see in fact that the entire statistical distribution of $n_1$ is the same as in the uniform trimer case.

\begin{figure}
\centerline{
\includegraphics[width=0.5\textwidth]{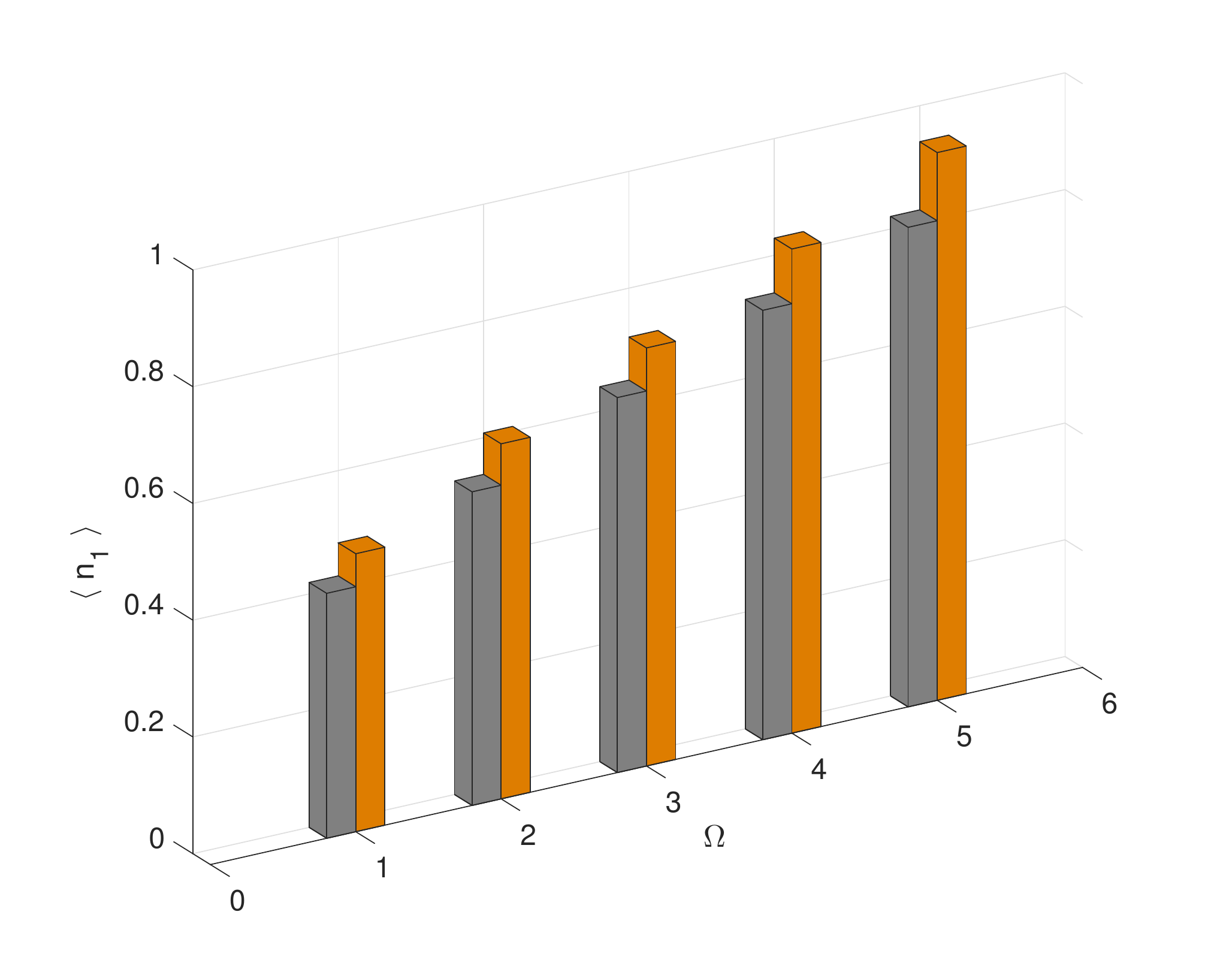}
}
\caption{\label{fig_disordered2}\textbf{Boson number expectation value on site 1 as function of varying drive strength ($\Omega$) for a disordered trimer.} Gray bars correspond to the lower sign choice in the parameters listed in Table~\ref{tab1}(b), Case 2; and orange bars correspond to the upper sign choice. The transformation from the lower sign choice to the upper sign choice is not number-conserving and the boson number expectation value is not invariant. }
\end{figure}

On the other hand, the parameter transformation in Case 2 is not of the type with boson number symmetry discussed earlier.  Consequently, Fig. (\ref{fig_disordered2}) shows that $\langle n_1\rangle$ is different between $|{\rho}^{(+)}_\infty\rangle$ and $|{\rho}^{(-)}_\infty\rangle$ at each value of $\Omega$.

Finally, we note that although we only consider the observables in the NESS, the invariance under the number-conserving transformation is at the level of  EOM, and the dynamical observables should also remain invariant. 

\section*{Discussion}
We have given an analytical argument and provided numerical evidence for a boson number symmetry of the DDBHM.  Specifically, the symmetry is that the boson number statistics of the system state are invariant to collective changes in the sign of the interaction energies, detunings, and hopping energies.  In other words, simultaneously changing the sign of all of the parameters of the number-conserving terms of the system Hamiltonian does not observably change the state.  On the other hand, we have also numerically shown that keeping the sign of the hopping energy fixed while changing the signs of the detunings and interaction energies does not leave the number statistics invariant.

We have therefore shown two contrasts to the case of equilibrium phases of the BHM: 1) the number statistics of the NESS of the DDBHM can exhibit a strong dependence on the sign of the hopping energy, and 2) it is possible for the number statistics of the NESS to be exactly the same for opposite signs of the interaction energy with the same magnitude.

These theoretical predictions are experimentally testable with existing superconducting circuit technology, and the symmetry is applicable beyond the DDBHM to any situation where the Lindbladian jump operators are real.

For self-interactions of bosons in superconducting circuits, it is strong \textit{attractive} interactions that are experimentally accessible \cite{bourassa2012josephson,hacohen2015cooling} rather than strong \textit{repulsive} interactions.  Therefore, the equivalence between attractive and repulsive interactions that we have shown here for driven-dissipative bosonic phenomena indicates that superconducting circuits with strong attractive interactions are a viable platform for investigating predictions made for driven-disspative bosonic phenomena involving strong repulsive interactions, such as repulsively induced photon superbunching \cite{grujic2013repulsively}, fermionized photons \cite{carusotto2009fermionized}, polariton crystalization \cite{hartmann2010polariton}, photon transport resonances \cite{biella2015photon}, first-order dissipative quantum phase transitions \cite{Weimer2015-is}, and diffusive-insulator transport phase transitions \cite{debnath2017nonequilibrium}.

\textit{Note added:} After our paper was written, we noticed a preprint by Li and Koch \cite{koch2017} reaching similar conclusions. 

\section*{Methods}
\subsection{Effective Equation of Motion}
The open boundary Bose-Hubbard chain under homogeneous coherent driving is described by the following Hamiltonian ($\hbar=1$):
\begin{align}
H=&\sum_l \omega_{l}b_l^{\dagger}b_l - \sum_l J_{l,l+1}(b_l^{\dagger}b_{l+1} + b_lb_{l+1}^{\dagger}) \nonumber \\&+ \sum_l \frac{U_l}{2}b_l^{\dagger}b_l^{\dagger}b_lb_l 
+ \sum_l(\Omega b_l^{\dagger}e^{-i\omega_{d}t}+\Omega^{*}b_le^{+i\omega_{d}t}).
\end{align}
$\omega_{l}$ denotes the bare frequency of the $l$th site, $J_{l,l+1}$ denotes the hopping amplitude between the $l$th and ($l+1$)th site, $U_l$ denotes the boson interaction energy on the $l$th site, $\Omega$ denotes the drive amplitude, and $\omega_{d}$ denotes the drive frequency.  With on-site dissipation to a Markovian bath, the density matrix $\rho$ of the chain is governed by the following Lindblad master equation: $\frac{d}{dt}\rho=\mathcal{L}\rho=-i[H,\rho] + \gamma\sum_l\mathcal{D}[b_l]\rho$, where $\mathcal{D}[b]\rho=\frac{1}{2}(2b\rho b^{\dagger}-b^{\dagger}b\rho-\rho b^{\dagger}b)$ and $\gamma$ is the local dissipation rate.  To eliminate the time dependence, the master equation is multiplied from the left by $U$ and from the right by $U^{\dagger}$, where $U=e^{i\omega_d t\sum_l  b_l^{\dagger}b_l}$.  The resulting effective EOM is (setting $\Omega$ real)
\begin{align}
\frac{d}{dt}\tilde{\rho}=&\mathcal{\tilde{L}}\tilde{\rho}\nonumber\\=&-i[\tilde{H},\tilde{\rho}] + \gamma\sum_l \frac{1}{2}(2b_l\tilde{\rho}b_l^{\dagger}-b_l^{\dagger}b_l\tilde{\rho}-\tilde{\rho}b_l^{\dagger}b_l),\\
\tilde{H}=& \sum_l \Delta_{l}b_l^{\dagger}b_l - \sum_l J_{l,l+1}(b_l^{\dagger}b_{l+1} + b_lb_{l+1}^{\dagger}) \nonumber\\
&+ \sum_l \frac{U_l}{2}b_l^{\dagger}b_l^{\dagger}b_lb_l + \Omega\sum_l(b_l^{\dagger}+b_l),
\end{align}
where $\tilde{\rho}=U\rho U^{\dagger}$ and $\Delta_{l}=\omega_l-\omega_d$ is the site-dependent drive detuning, which plays the role of a chemical potential.  For simplicity, in the main text we write $\tilde{\rho}$ as $\rho$ and $\tilde{H}$ as $H$.

\subsection{Numerical Simulations}
The numerical simulation is performed by employing a matrix product density operator (MPDO) representation of ${\rho}$ \cite{Zwolak2004-tz, Verstraete2004-hc}, which amounts to a quantum mechanical treatment characterized by a refinement parameter $\chi$ that designates the maximum size of the tensors that represent each site, and therefore the maximum amount of total correlations (classical plus quantum) between bipartitions of the chain that can be captured by the MPDO.  Linking each site tensor with its neighbor in the MPDO is a diagonal matrix of $\chi$ ``singular values" that represents these correlations.

In the MPDO picture the system density matrix ${\rho}$ becomes a vector, denoted $|{\rho}\rangle$, and the superoperator $\mathcal{{L}}$ becomes a regular operator $\mathcal{{L}}_\sharp$ such that $\langle {\rho}|\mathcal{{L}}_\sharp |{\rho}\rangle = 0$ at the NESS. To obtain an approximation for ${\rho}_{\infty}$ under a given set of system parameters $U_l$, $J_l$, $\Delta_l$, $\Omega$, and $\gamma$, we first use the hybrid evolution method of Ref. \cite{gangat2016steady} to evolve the MPDO representation of a random initial state ${\rho}$ under a desired choice of parameters until convergence in achieved.  We then sweep the value of $\Omega$ in increments, converging the MPDO with real time evolution at each increment.  Convergence is considered complete when $\langle \mathcal{{L}}_\sharp \rangle \lesssim 10^{-3}$ and the singular values between the first two sites of the MPDO are converged on a logarithmic scale.  We find that $\chi=15$ and a timestep size of $10^{-1}$ is sufficient to achieve this for all of the cases that we consider.  We verify uniqueness of the NESS by performing the sweep of $\Omega$ in both directions.  We truncate the Hilbert space on each site at four quanta, and always choose $\gamma=1$.
\section*{Acknowledgements}

This work is partially supported by Ministry of Science and Technology, Taiwan, under Grants No. MOST 104-2112-M-002 -022 -MY3,  MOST 105-2112-M-002-023-MY3, MOST 106-2811-M-002-054 (A.A.G., Y.J.K.). 

\section*{Author contribution statement}
A.A.G proposed the project, performed the numerical simulations and wrote the manuscript; I.P.M. proposed the general proof; Y.J.K. supervised the project and wrote the manuscript. All the authors discussed the results and the manuscript.
\section*{Additional information}

The authors declare that they have no competing financial interests. Correspondence and requests for materials should be addressed to Y.J.K. (email: yjkao@phys.ntu.edu.tw) .

\bibliographystyle{naturemag-doi}
\bibliography{Ref}

\end{document}